\documentclass[twocolumn,english,pra,showpacs,twocolumn]{revtex4}
\usepackage[T1]{fontenc}
\usepackage[latin9]{inputenc}
\usepackage{amsmath}
\usepackage{amssymb}
\usepackage{graphicx}
\usepackage{esint}

\makeatletter
\@ifundefined{textcolor}{}
{%
 \definecolor{BLACK}{gray}{0}
 \definecolor{WHITE}{gray}{1}
 \definecolor{RED}{rgb}{1,0,0}
 \definecolor{GREEN}{rgb}{0,1,0}
 \definecolor{BLUE}{rgb}{0,0,1}
 \definecolor{CYAN}{cmyk}{1,0,0,0}
 \definecolor{MAGENTA}{cmyk}{0,1,0,0}
 \definecolor{YELLOW}{cmyk}{0,0,1,0}
}

\usepackage{babel}

\makeatother

\usepackage{babel}
\begin{document}

\title{Indirect driving of cavity QED system and its induced non-linearity}

\author{Yusuf Turek$^{1,3}$, L. P.Yang$^{1,3}$, W. Maimaiti$^{1}$, Yong
Li$^{2,3}$,}

\email{liyong@csrc.ac.cn}

\selectlanguage{english}%

\author{C. P. Sun$^{2,3}$}

\email{cpsun@csrc.ac.cn}

\homepage{http://www.csrc.ac.cn/ suncp}

\selectlanguage{english}%

\affiliation{$^{1}$State Key Laboratory of Theoretical Physics, Institute of
Theoretical Physics, Chinese Academy of Sciences, and University of
the Chinese Academy of Sciences, Beijing 100190, China}

\affiliation{$^{2}$Beijing Computational Science Research Center, Beijing 100084,
China}

\affiliation{$^{3}$Synergetic Innovation Center of Quantum Information and Quantum
Physics, University of Science and Technology of China, Hefei, Anhui
230026, China}
\begin{abstract}
The linear driving for a single-mode optical field in a cavity can
result from the external driving of classical field even when the
coupling between the classical field and the cavity is weak. We revisit
this well known effect with a microscopic model where a classical
field is applied to a wall of the cavity to excite the atoms in the
wall, and re-combination of the low excitations of the wall mediates
a linear driving for the single-mode field inside the cavity. With
such modeling about the indirect driving through the quantum excitations
of the wall, we theoretically predict several non-linear optical effects
for the strong coupling cases, such as photon anti-bunching and photon
squeezing. In the sense, we propose the most simplified non-linear
quantum photonics model.
\end{abstract}


\date{\today}

\maketitle

\section{Introduction}

Photons are prior candidate for quantum information processing such
as quantum computing and long distance quantum communication, as they
can be easily generated and can travel long distances with high coherence.
Due to the ability of obtaining photon-photon interactions, the nonlinear
optical process has great advantages in quantum information processing
and quantum computation compared with linear optics methods, and possesses
great potential for a variety of emerging technologies~\cite{Boyd}.
However, there is no direct interaction between single photons in
physics according to the quantum electrodynamics (QED). Hence, it
is of great importance to achieve the interaction between single photons.
The most popular method to generate the strong nonlinear effects between
photons is spontaneous parametric down-conversion, which is used especially
as a source of entangled photon pairs~\cite{Klyshko,Burnham}. Examples
of such quantum optical phenomena have been investigated experimentally
including generation of quadrature squeezing states and two-photon
entanglement sates in various degrees of freedom~\cite{Walls,Wu,Loudon,Kwiat,Brendel,Mair,Ramelow,Shih,Barreino}.

Generally, the typical nonlinear optical phenomenon occurs only at
very high optical intensities and the degree of nonlinearity between
single photons is very low~\cite{Roussev,Vandevender,Thew}. However,
producing high-degree nonlinearity at very low mean-photon level is
desirable in many quantum information processing applications~\cite{Cirac,Imamoglu},
such as, photon blockade effect which plays an important role as an
effective single photon source in quantum information processing.
Recently, some similar nonlinear effects such as cross phase modulation
\cite{Matsuda,Lo} and spontaneous downconversion \cite{Hubel,Shalm}
have been observed with a single-photon level pump. These nonlinearities
at single-photon level are also obtained through optical cavity in
which photon-photon interaction is relatively strong \cite{Walls}.
Recently, Gupta \textit{et al.} \cite{Gupta } experimentally investigated
the Kerr nonlinearity and dispersive optical bistability of a Fabry-Perot
(FP) optical cavity arising from the long lived coherent motion of
ultracold atoms trapped within. They reported that the strong nonlinearity
would be observed at low average intracavity photon number level $\bar{n}=0.05$,
and even at as low as $\bar{n}=10^{-4}$. The photon blockade effect
was also found in optomechanical systems~\cite{Rabl}, where the
Kerr interaction between photons is induced by the strong optomechanical
coupling.

It is known that the linear coupling between an external classical
field and a single-mode cavity can result in a linear driving to create
a coherent state of the cavity field. If we assume that the single-mode
cavity field is driven by an external driving field with frequency
$\omega_{f}$, then the Hamiltonian can be written as
\[
V_{f}=\omega a^{\dagger}a+f_{0}a^{\dagger}e^{-i\omega_{f}t}+H.c.,
\]
where $a^{\dagger}$$\left(a\right)$ is the creation (annihilation)
operator of the single-mode radiation field and $f_{0}$ is the related
driving strength. The corresponding energy spectrum of the output
field of the cavity is of the Lorentz form. However, the underlying
mechanism to explain this simple phenomenon is not clear until now.
In this paper, we revisit this well known effect by giving a microscopic
explanation of physical mechanism for such linear driving. Here, a
classical field is applied to one wall of a FP cavity, which is modeled
as a two-level atomic ensemble where the two levels can be imagined
as excited and non-excited states of local exictons. When the decay
rate of the atomic ensemble is much larger than the decay of the cavity,
the re-combination of the low excitations of the ensemble in the wall
will mediate a linear driving for the single-mode field inside the
cavity. Furthermore, if the higher-order excitation of the atomic
ensemble is taken into account, the single-mode cavity field exhibits
some interesting nonlinear photonic phenomena.

When the external driving field is weak, but the coupling between
atomic ensemble (the cavity wall) and the cavity mode is strong, the
Kerr non-linear effect is dominant in the effective Hamiltonian of
photons, which produces the photon blockade phenomena. In this case,
we found that the strong nonlinearity as well as photon blockade of
our system would occurr at low intracavity photon number, even tough
as low as $\bar{n}\simeq10^{-4}$.

On the contrary, for the case of strong driving and weak coupling
(between the atomic ensemble and the cavity), the light-squeezing
nonlinear effect is dominant. For weak coupling between atomic ensemble
and single-mode cavity field we found that in our system the optical
bistability, even quad-stability phenomena would appear with increasing
the driving strength. From the output intensity spectrum of single-mode
cavity field, we found that the squeezed effect of the output field
occurs when the driving strength increases. We also found that the
maximum squeezing takes place at the vicinity of resonance point.

This paper is organized as follows: In Sec.~II, we describe our model
with an effective Hamiltonian in terms of collective low excitation
operators of atomic ensemble (cavity wall), and present the clear
microscopic explanation to indirect quantum driving. In Sec. III,
we consider the effects of higher-order excitation of atomic ensemble
(cavity wall) and obtain the effective Hamiltonian of single-mode
cavity field, which describes the very interesting nonlinear photonic
phenomena by controlling some corresponding parameters of the system.
In Sec. IV, we study the two extreme cases separately, and calculate
the second-order correlation function and output spectrum. Finally,
we make conclusion and give some remarks to our work in Sec.V.

\section{Simplified model for nonlinear photonics and its linear limit}

\begin{figure}
\includegraphics[width=8cm]{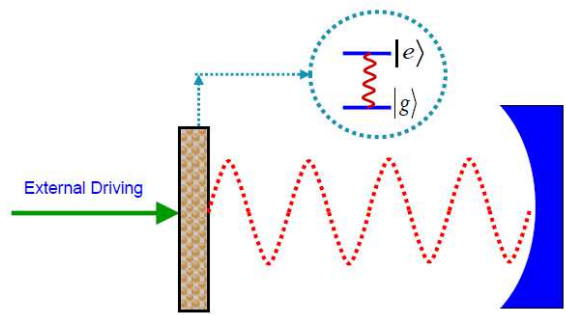}

\caption{(Color online) Schematic of indirect quantum driving model. The wall
of the cavity consists of $N$ two-level atoms with the same energy
difference $\omega_{a}$. A classical field with frequency $\omega_{f}$
is applied to excite these atoms to generate a linear driving for
the single-mode cavity field.}

\label{model-1}
\end{figure}

In this section, we build a microscopic model to explain the quantum
driving. Here, we assume that a wall (left wall) of the cavity consists
of a vast amount of two-level systems (TLSs), which can be viewed
as an atomic ensemble. The two levels can be imagined as the excited
and non-excited sates of local exciton. As shown in Fig.~\ref{model-1},
the left wall of the cavity is driven by a classical external field
with frequency $\omega_{f}$. The model Hamiltonian reads as (hereafter
we take $\hbar=1$)
\begin{eqnarray}
\!\!\!\!\!\! H & \!\!= & \!\!\omega_{c}c^{\dagger}c\!+\!\sum_{i=1}^{N}\!\{\frac{\omega_{a}}{2}\sigma_{z}^{\left(i\right)}\!+\!\![(gc\!+\!\Omega e^{-i\omega_{f}t})\sigma_{+}^{\left(i\right)}\!+\!\text{H.c.}]\}\!,\label{eq:model Hamil}
\end{eqnarray}
where, $c\ (c^{\dagger})$ is the annihilation (creation) operator
of single-mode cavity field with frequency $\omega_{c}$, the Pauli
matrices $\sigma_{z}^{\left(i\right)}=\left|e_{i}\right\rangle \left\langle e_{i}\right|-\left|g_{i}\right\rangle \left\langle g_{i}\right|$,
$\sigma_{+}^{\left(i\right)}=\left|e_{i}\right\rangle \left\langle g_{i}\right|$,
and $\sigma_{-}^{\left(i\right)}=\left|g_{i}\right\rangle \left\langle e_{i}\right|$
describe the $i$-th atom with the ground (excited) states $\left|g_{i}\right\rangle $
($\left|e_{i}\right\rangle $) and energy level spacing $\omega_{a}$;
$N$ is the number of the two-level atoms, and $\omega_{f}$ is the
frequency of the classical driving field. For simplicity, we take
the uniform driving strength $\Omega_{i}=\Omega$ and cavity-atom
coupling constant $g_{i}=g$.

To explore the novel effects and phenomena resulting from above, we
take the Holstein-Primakoff (H-P) transformation~\cite{Holstein}
for the collective atomic operators
\begin{eqnarray}
\sum_{i=1}^{N}\sigma_{+}^{\left(i\right)} & = & B^{\dagger}\sqrt{N-B^{\dagger}B},\label{eq:up opera}\\
\sum_{i=1}^{N}\sigma_{-}^{\left(i\right)} & = & \sqrt{N-B^{\dagger}B}B,\label{eq:down opera}
\end{eqnarray}
 and
\begin{equation}
\sum_{i=1}^{N}\sigma_{z}^{\left(i\right)}\equiv2B^{\dagger}B-N.\label{eq:low sigmaz}
\end{equation}
 Here, the $B$ and $B^{\dagger}$ represent the atomic collective
excitation operators. In the low excitation limit $\langle B^{\dagger}B\rangle/N\ll1$,
we have ~\cite{Song,JIn,sun-1}

\begin{equation}
B^{\dagger}\approx\frac{1}{\sqrt{N}}\sum_{i=1}^{N}\sigma_{+}^{\left(i\right)},\ {\rm and}\text{ }B\approx\frac{1}{\sqrt{N}}\sum_{i=1}^{N}\sigma_{-}^{\left(i\right)},\label{eq:B boson-1}
\end{equation}
where the operator $B$ satisfies the standard bosonic commutation
relation $\left[B,B^{\dagger}\right]\approx1.$ Using these relations
(\ref{eq:low sigmaz}) and (\ref{eq:B boson-1}), in the interaction
picture with respect to $H_{0}=\omega_{f}\left(c^{\dagger}c+B^{\dagger}B\right)$
we can rewrite our model Hamiltonian (\ref{eq:model Hamil}) in terms
of the atomic collective operators $B$ and $B^{\dagger}$ as
\begin{equation}
H^{\left(0\right)}=\Delta_{c}c^{\dagger}c+\Delta_{b}B^{\dagger}B+\left(GcB^{\dagger}+\chi B+\text{H.c.}\right),\label{eq:inteac pic}
\end{equation}
where $\Delta_{c}=\omega_{c}-\omega_{f}$ is the detuning between
the single-mode cavity and external driving field and $\Delta_{b}=\omega_{a}-\omega_{f}$
the detuning between two-level atom and external field, $G=g\sqrt{N}$
and $\chi=\Omega\sqrt{N}$. For simplicity here, we assumed all these
coupling strengths are real. We note that during the derivation of
Eq. (\ref{eq:inteac pic}) we have neglected a constant term $N\omega_{b}/2$
since it has no effect on our results in the context.

The quantum Langevin equations of variables of our system are obtained
from Eq. (\ref{eq:inteac pic}) as
\begin{equation}
\dot{c}\left(t\right)=-i\Delta_{c}c\left(t\right)-iGB\left(t\right)-\frac{\kappa}{2}c\left(t\right)+\sqrt{\kappa}c_{in}\left(t\right),\label{eq:langiven c}
\end{equation}

\begin{equation}
\dot{B}\left(t\right)\!=\!-i\Delta_{b}\! B\left(t\right)\!-\! iGc\left(t\right)\!-\! i\chi\!-\!\frac{\gamma}{2}B\left(t\right)\!+\!\sqrt{\gamma}B_{in}\left(t\right)\!,\label{eq:Langiven B}
\end{equation}
where $\kappa$ is the decay rate of the cavity, $\gamma$ is the
decay rate of collective mode $B$, and $c_{in}\left(t\right)$ and
$B_{in}\left(t\right)$ are zero-meannoise operators (i.e., $\langle c_{in}\rangle=\langle B_{in}\rangle=0$)
satisfying the fluctuation relations \begin{subequations}

\begin{eqnarray}
\langle c_{\mathrm{in}}(t)c_{\mathrm{in}}^{\dagger}(t^{\prime})\rangle & = & [n(\omega_{c})+1]\delta(t-t^{\prime}),\label{eq: nfluc of c}\\
\langle B_{\mathrm{in}}(t)B_{\mathrm{in}}^{\dagger}(t^{\prime})\rangle & = & [n(\omega_{b})+1]\delta(t-t^{\prime}),\label{eq:nfluc of B}
\end{eqnarray}
\end{subequations}where

\begin{equation}
n(\omega_{r})=\frac{1}{\exp\left(\frac{\omega_{r}}{k_{B}T}\right)-1},\ \ (r=b,c)\label{eq:thermal num}
\end{equation}
are the average thermal excitation numbers of the cavity mode and
atomic collective modes at temperature $T$, respectively.

In our system, the natural life time of exited atom is much smaller
than the life time of a photon in the cavity, $\gamma^{-1}\ll\kappa^{-1}$.
Thus we can eliminate adiabatically the degrees of freedom of the
atomic ensemble by substituting the steady-state solution of Eq.~(\ref{eq:Langiven B})
into Eq. (\ref{eq:langiven c}) and obtain
\begin{equation}
\dot{c}\left(t\right)=-i\Delta_{eff}^{\left(0\right)}\left(t\right)-\frac{1}{2}\kappa c\left(t\right)-if+\sqrt{\kappa}c_{in}\left(t\right).\label{eq:eff Langiven}
\end{equation}
Here, $\Delta_{eff}^{\left(0\right)}=\omega_{eff}^{\left(0\right)}-\omega_{f}$
is the detuning between the driving field and the effective frequency
of single mode cavity $\omega_{eff}^{\left(0\right)}=\omega_{c}-\delta$
with the atomic ensemble induced shift $\delta=Ng^{2}/\omega_{ba}$
and $\omega_{ac}=\omega_{a}-\omega_{c}$, and $f=-G\chi/\Delta_{b}$
is the induced driving amplitude of single-mode cavity. Here, we assume
that the laser is detuned sufficiently far from resonance that $\vert\Delta_{b}\vert\gg\gamma,\ \Omega$.
Under this condition, the driven atomic ensemble only modifies the
resonance frequency of the cavity, and the correction of the cavity
decay rate has been neglected.

From Eq. (\ref{eq:eff Langiven}), we obtain an effective Hamiltonian
of single-mode cavity field as

\begin{eqnarray}
H_{eff} & = & \omega_{eff}^{\left(0\right)}c^{\dagger}c+(fc^{\dagger}e^{-i\omega_{f}t}+{\rm H.c}.),\label{eq:Eff Hamil of c}
\end{eqnarray}
 which describes a typical model of a quantum harmonic oscillator
driven by a classical field with strength $f$ and frequency $\omega_{f}$.

\begin{figure}
\includegraphics[width=8cm]{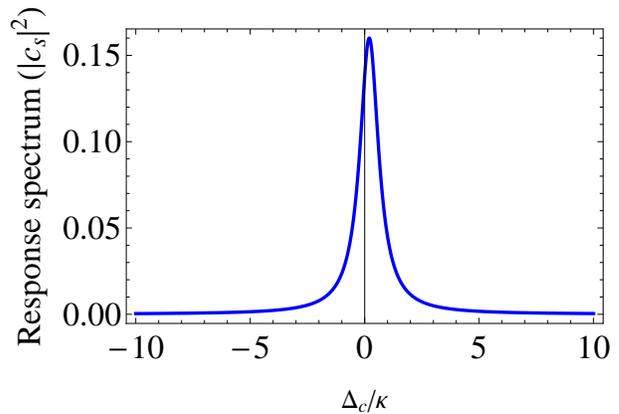}

\caption{\label{fig:1}(Color online) Cavity response vs detuning. Here, $\omega_{ac}=\omega_{a}-\omega_{c}=5\times10^{4}$,
$N=1\times10^{4}$, $\Omega=1$, $g=1$. All parameters are in the
units of cavity decay rate, $\kappa$. In this figure the shift amount
is $\delta=0.2$.}
\end{figure}

As shown in Fig. (\ref{fig:1}), the driving on the wall of the cavity
induces an effective driving for the cavity mode and the spectrum
of the output field is of the Lorenz form centered at the effective
frequency $\omega_{eff}^{(0)}$. This is the so-called indirect driving.

\section{Nonlinear photonic effect due to larger excitation}

When the low excitation condition breaks, we need to take the higher
order of the H-P transformation into account,
\begin{align}
\sum_{i=1}^{N}\sigma_{+}^{\left(i\right)} & \approx\sqrt{N}B^{\dagger}\left(1-\frac{B^{\dagger}B}{2N}\right),\label{eq:first exp of up sig}\\
\sum_{i=1}^{N}\sigma_{-}^{\left(i\right)} & \approx\sqrt{N}\left(1-\frac{B^{\dagger}B}{2N}\right)B,\label{eq:first exp of down sig}
\end{align}
 and
\begin{equation}
S_{z}=\sum_{i=1}^{N}\sigma_{z}^{\left(i\right)}=B^{\dagger}B-\frac{N}{2}.\label{eq:singma z}
\end{equation}
For this case, the Hamiltonian (\ref{eq:inteac pic}) is rewritten
as
\begin{align}
H' & =H^{\left(0\right)}+H^{\left(1\right)},\label{eq:fist order Hamli}
\end{align}
where the first part is the zeroth order form as given in Eq.~(\ref{eq:inteac pic})
and the second part is
\begin{equation}
H^{\left(1\right)}=-\frac{1}{2N}\left(GcB^{\dagger2}B+\chi B^{\dagger2}B+\text{H.c.}\right),\label{eq:first order Hamil}
\end{equation}
resulting from the first order expansion of collective excitation
operator. We can see that, for the low excitation case $\langle B^{\dagger}B\rangle/N\ll1$,
the effect of $H^{\left(1\right)}$can be neglected, and the Hamiltonian
(\ref{eq:fist order Hamli}) reduces to the zeroth order Hamiltonian
(\ref{eq:inteac pic}).

We note that the adiabatic elimination does not depend on the number
of the atoms in the ensemble \cite{Gardiner-1}. Thus here, we can
still use the adiabatic elimination method to get the effective Hamiltonian
of single-mode cavity field. By taking the same procedure as in Sec.
II, we obtain the effective Hamiltonian of the single mode cavity
field as

\begin{eqnarray}
\!\!\!\!\!\!\!\!\!\!\!\!\!\tilde{H}_{eff}\!\!\! & = & \!\!\!\triangle_{eff}^{\left(1\right)}c^{\dagger}c\!+\!\!\chi_{{\rm kerr}}c^{\dagger2}c^{2}\!\!+\!\!\left[\!\mu c^{2}\!\!+\!\zeta c^{\dagger}c^{2}\!+\! Fc\!+\!\text{H.c.}\right]\!\!.\label{eq:effec first order}
\end{eqnarray}
Here,\begin{subequations}
\begin{align}
\Delta_{eff}^{\left(1\right)} & =\Delta_{c}-\frac{Ng^{2}}{\Delta_{b}}+\chi_{{\rm kerr}}+4\mu,\label{eq:first order eff det}\\
\chi_{{\rm kerr}} & =\frac{Ng^{4}}{\Delta_{b}^{3}},\label{eq:kerr coeff}\\
\mu & =\frac{Ng^{2}\Omega^{2}}{\Delta_{b}^{3}},\label{eq:squeezing coeff}\\
\zeta & =\frac{2Ng^{3}\Omega}{\Delta_{b}^{3}},\label{eq:phase match}
\end{align}
and\end{subequations}
\begin{equation}
F=-\frac{Ng\Omega}{\Delta_{b}}\left[1-\frac{2\Omega^{2}+g^{2}}{\Delta_{b}^{2}}\right].\label{eq:efffec ampiltu}
\end{equation}
From this result we can see that if we take the the first order expansion
of the collective atomic operators, the effects of photonic nonlinearity
appear. Here, the term $c^{\dagger2}c^{2}$ characterizes the Kerr
effect with the strength $\chi_{kerr}$, $c^{2}$ charaterizes the
squeezing effect with the strength $\mu$, and $c^{\dagger}c^{2}$
denotes the two photon phase-space filling effect with the strength
$\zeta$. We note that if $\vert\Delta_{b}\vert\gg\sqrt{N}g,\Omega$
the effects of these nonlinear terms are negligible. We also note
that the strengths of these terms can be controlled and enhanced separately
by mediating the corresponding parameters.

If the atom-cavity coupling strength is much larger than that of the
external driving field, i.e., $g\gg\Omega$, the above effective Hamiltonian
(\ref{eq:effec first order}) of single-mode cavity field reduces
to
\begin{equation}
H_{1}=\Delta_{eff,1}c^{\dagger}c+\chi_{{\rm kerr}}c^{\dagger2}c^{2}+(F^{\prime}c+\zeta c^{\dagger}c^{2}+\text{H.c.}),\label{eq:Kerr Hamil}
\end{equation}
where we have neglected the squeezing term, and
\begin{equation}
\Delta_{eff,1}=\Delta_{c}-\frac{Ng^{2}}{\Delta_{b}}+\chi_{{\rm kerr}},\label{eq:Kerr detu}
\end{equation}
 and
\begin{equation}
F^{\prime}=-\frac{Ng\Omega}{\Delta_{b}}+\frac{1}{2}\zeta,\label{eq:Kerr dri ampli}
\end{equation}
are the corrected detuning and driving strength, respectively. The
strength of the Kerr term $\chi_{{\rm kerr}}$ is related to the number
of the atoms $N$, the atomic detuning $\Delta_{b}$, and the atom-cavity
coupling strength $g$, but independent of $\Omega$. Thus, we can
enhance the Kerr term effect by mediating $g$, $\Delta_{b}$ with
fixed number of the atoms. In this case, we can investigate the photon
statistical properties of single-mode cavity field.

If the strength of external driving field is much larger than the
atom-cavity coupling strength, i.e., $\Omega\gg g$, the total effective
Hamiltonian (\ref{eq:effec first order}) reduces to
\begin{equation}
H_{2}=\Delta_{eff,2}c^{\dagger}c+(F^{\prime\prime}c+\mu c^{2}+\zeta c^{\dagger}c^{2}+\text{H.c.}).\label{eq:Squeezing Hamil}
\end{equation}
Here,
\begin{equation}
\Delta_{eff,2}=\Delta_{c}-\frac{Ng^{2}}{\Delta_{b}}+4\mu,\label{eq:Squeezing detu}
\end{equation}
and
\begin{equation}
F^{\prime\prime}=-\frac{Ng\Omega}{\Delta_{b}}+\frac{2Ng\Omega^{3}}{\Delta_{b}^{3}}.\label{eq:Squeezing amplitu}
\end{equation}
We can see that in this particular case the dominant squeezing effect
of light and other correlated photonic nonlinear effects can be directly
controlled by the external driving strength. In this case, we can
calculate the output squeezing spectrum of the cavity field to investigate
the efficiency of our scheme to generate the squeezed photonic state.

Note that in both the cases discussed above we have to consider the
two-photonic phase space filling effect term, whose strength is characterized
by $\zeta$. Since its strength is related to $g$ and $\Omega$,
it will directly effect the investigated both nonlinear phenomena.
In the next, we will investigate the above two particular cases separately.

\section{Second order correlation - Photon antibunching}

In this section, we study the first case $g\gg\Omega$, where the
Kerr effect is dominant. To investigate the photon statistics of single-mode
cavity radiation field, we will calculate the second order correlation
function at zero time delay, $g^{\left(2\right)}\left(0\right)=\langle a^{\dagger}a^{\dagger}aa\rangle/\langle a^{\dagger}a\rangle^{2}$.
We will begin our calculation by writing the master equation of our
system with Hamiltonian (\ref{eq:Kerr Hamil})
\begin{align}
\dot{\rho} & =-i\left[H_{1},\rho\right]+\kappa\left(n_{th}+1\right)\left(2c\rho c^{\dagger}-c^{\dagger}c\rho-\rho c^{\dagger}c\right)\nonumber \\
 & +\kappa n_{th}\left(2c^{\dagger}\rho c-cc^{\dagger}\rho-\rho cc^{\dagger}\right),\label{eq:master equ}
\end{align}
where, $n_{th}=n\left(\omega_{c}\right)$ is the thermal occupation
number of the single-mode cavity field as defined in (\ref{eq:thermal num}).
As we see, in our system the operator equation is nonlinear, in this
case it is useful to use the $c$-number Fock-Planck equation.

The density matrix of the cavity mode in generalized $P$ representation
function\cite{Walls} reads
\begin{equation}
\rho=\int\Lambda\left(\boldsymbol{\alpha}\right)P\left(\alpha,\beta\right)d\mu\left(\alpha,\beta\right),
\end{equation}
where $\left(\boldsymbol{\alpha}\right)=\left(\alpha,\beta\right)\equiv\left(\alpha,\alpha^{\dagger}\right)$,
and in generalized $P$ representation $\alpha$ and $\alpha^{\dagger}$
are independent variables. The non-diagonal coherent state projection
operator is defined as
\begin{equation}
\Lambda\left(\boldsymbol{\alpha}\right)=\frac{\left|\alpha\right\rangle \left\langle \beta^{\ast}\right|}{\left\langle \beta^{\ast}\right|\left.\alpha\right\rangle }.
\end{equation}
The corresponding Fock-Plank equation of $\rho$ in the $P$ representation
is written as \begin{widetext}

\begin{align}
\frac{\partial P\left(\alpha\right)}{\partial t} & =\frac{\partial}{\partial\alpha}\left[\kappa^{\prime}\alpha+\zeta^{\prime}\left(\alpha^{2}+2\alpha^{*}\alpha\right)+2\chi''\alpha^{*}\alpha^{2}-E\right]P+\frac{\partial}{\partial\alpha^{*}}\left[\kappa{}^{\prime\ast}\alpha^{*}+\zeta{}^{\prime\ast}\left(\alpha^{*2}+2\alpha^{*}\alpha\right)+2\chi^{\prime\prime}{}^{\ast}\alpha\alpha^{*2}-E^{\ast}\right]P\nonumber \\
 & -\frac{\partial^{2}}{\partial\alpha^{2}}\left[\chi^{\prime\prime}\alpha^{2}+\zeta^{\prime\prime}\alpha\right]P-\frac{\partial^{2}}{\partial\alpha^{*2}}\left[\chi^{\prime\prime\ast}\alpha^{*2}+\zeta{}^{\prime\ast}\alpha^{\ast}\right]P+2\kappa n_{th}\frac{\partial^{2}}{\partial\alpha\partial\alpha^{\ast}}P.\label{eq:F-P Eq}
\end{align}
\end{widetext}Here, $\kappa^{\prime}=\kappa+i\Delta_{eff,1}$, $\chi^{\prime\prime}=i\chi{}_{kerr}$,
$\zeta^{\prime}=i\zeta$, and $E=-iF^{\prime}$. In the $P$ representation,
$\alpha$ and $\alpha^{\ast}$ are independent complex variables and
the Fokker-Planck equation has a positive semi-definite diffusion
matrix in four-dimensional space. This allows us to define the equivalent
stochastic differential equations using the Ito rules \cite{Walls}\begin{widetext}
\begin{align}
\frac{\partial}{\partial t}\left(\begin{array}{c}
\alpha\\
\alpha^{\ast}
\end{array}\right)= & \left(\begin{array}{c}
E-\kappa^{\prime}\alpha-2\chi^{\prime\prime}\alpha^{\dagger}\alpha^{2}-\zeta^{\prime}\left(\alpha^{2}+2\alpha^{\dagger}\alpha\right)\\
E^{\ast}-\kappa{}^{\prime\ast}\alpha^{\dagger}-2\chi^{\prime\prime}{}^{\ast}\alpha^{\dagger2}\alpha-\zeta{}^{\prime\ast}\left(\alpha^{\dagger2}+2\alpha^{\dagger}\alpha\right)
\end{array}\right)+\left(\begin{array}{cc}
-2\chi''\alpha^{2}-2\zeta^{\prime}\alpha & 2\kappa n_{th}\\
2\kappa n_{th} & -2\chi^{\prime\prime}{}^{\ast}\alpha^{\dagger2}-2\zeta{}^{\prime\ast}\alpha^{\dagger}
\end{array}\right)^{\frac{1}{2}}\left(\begin{array}{c}
\eta_{1}\left(t\right)\\
\eta_{1}^{\dagger}\left(t\right)
\end{array}\right),\label{eq:SDE}
\end{align}
\end{widetext}where $\eta_{1}\left(t\right)$ and $\eta_{1}^{\dagger}\left(t\right)$
are the delta correlated stochastic forces with zero mean, namely\begin{subequations}
\begin{eqnarray}
\langle\eta_{1}\left(t\right)\rangle & = & \langle\eta_{1}^{\dagger}\left(t\right)\rangle=0.\\
\langle\eta_{1}\left(t\right)\eta_{1}^{\dagger}\left(t'\right)\rangle & = & \delta\left(t-t'\right),\\
\langle\eta_{1}\left(t\right)\eta_{1}\left(t'\right)\rangle & = & 0,
\end{eqnarray}
\end{subequations}The semi-classical or mean value of the above equations
can obtained by replacing $\alpha^{\dagger}$ by the steady state
value $\alpha_{0}^{\ast}$ determined by
\begin{equation}
E-\kappa^{\prime}\alpha-2\chi^{\prime\prime}\alpha^{\dagger}\alpha^{2}-\zeta^{\prime}\left(\alpha^{2}+2\alpha^{\dagger}\alpha\right)=0,
\end{equation}
where $\vert\alpha_{0}\vert^{2}=n_{0}$ represents the mean intracavity
photon number in the steady state.

To investigate the effect of quantum fluctuation to steady state,
we consider the very small fluctuation around the steady state by
taking $\alpha=\alpha_{0}+\alpha_{1}$. Linearizing Eq. (\ref{eq:SDE}),
we obtain the stochastic differential equation for the fluctuation
variable $\alpha_{1}$$\left(\alpha_{1}^{\dagger}\right)$ as~\cite{Walls}

\begin{equation}
\frac{\partial}{\partial t}\boldsymbol{\alpha_{1}\left(t\right)=-A}.\boldsymbol{\alpha_{1}\left(t\right)}+\boldsymbol{D}^{\frac{1}{2}}.\boldsymbol{\xi}\left(t\right).
\end{equation}
Here, $\boldsymbol{\alpha_{1}}=\left(\alpha_{1},\alpha_{1}^{\dagger}\right)^{T}$,
and
\begin{equation}
\!\!\!\boldsymbol{A}\!\!=\!\!\left(\!\!\!\!\begin{array}{cc}
\kappa^{\prime}+4\chi^{\prime\prime}n_{0}\!+\!4\zeta^{\prime}\Re\left(\alpha_{0}\right), & \!\!\!\!\!\!\!\!\!\!2\chi^{\prime\prime}\alpha_{0}^{2}+2\zeta^{\prime}\alpha_{0}\\
2\chi^{\prime\prime}{}^{\ast}\alpha_{0}^{\ast2}+2\zeta{}^{\prime\ast}\alpha_{0}^{\ast}, & \!\!\!\!\!\!\!\!\kappa{}^{\prime\ast}\!+\!4\chi^{\prime\prime}{}^{\ast}n_{0}\!+\!4\zeta{}^{\prime\ast}\Re\left(\alpha_{0}\right)
\end{array}\!\!\!\!\right)\!\!,\label{eq:drift mat}
\end{equation}
 represents the drift matrix,

\begin{equation}
\boldsymbol{D}=\left(\begin{array}{cc}
-2\chi^{\prime\prime}\alpha_{0}^{2}-2\zeta^{\prime}\alpha_{0} & 2\kappa n_{th}\\
2\kappa n_{th} & -2\chi^{\prime\prime}{}^{\ast}\alpha_{0}^{\ast2}-2\zeta{}^{\prime\ast}\alpha_{0}^{\ast}
\end{array}\right)\label{eq:deffu mat}
\end{equation}
 is the diffusion matrix, and $\boldsymbol{\xi}\left(t\right)=\left(\eta_{1}\left(t\right),\eta_{2}\left(t\right)\right)^{T}$
.

According to Ref. \cite{Chaturvedi}, the correlation matrices

\begin{eqnarray}
\boldsymbol{C}_{ss} & \equiv & \left(\begin{array}{cc}
\langle a^{2}\rangle-\langle a\rangle^{2}, & \langle a^{\dagger}a\rangle-\vert\langle a\rangle\vert^{2}\\
\langle a^{\dagger}a\rangle-\vert\langle a\rangle\vert^{2} & \langle a^{\dagger2}\rangle-\langle a^{\dagger}\rangle^{2}
\end{array}\right)\nonumber \\
 & \approx & \left(\begin{array}{cc}
\langle\alpha_{1}^{2}\rangle, & \langle\alpha_{1}^{\dagger}\alpha_{1}\rangle\\
\langle\alpha_{1}^{\dagger}\alpha_{1}\rangle, & \langle\alpha_{1}^{\dagger2}\rangle
\end{array}\right)\label{eq:corre matrix}
\end{eqnarray}
in the steady state can be evaluated by
\begin{eqnarray}
\!\!\!\!\!\!\!\boldsymbol{C}_{ss} & \!\!= & \!\!\left(\begin{array}{cc}
C_{11} & C_{12}\\
C_{12} & C_{11}^{\ast}
\end{array}\right)\\
 & \!\!= & \!\!\frac{\boldsymbol{D}Det\left(\boldsymbol{A}\right)\!+\![\boldsymbol{A}\!-\! Tr\left(\boldsymbol{A}\right)I]\boldsymbol{D}[\boldsymbol{A}\!-\! Tr\left(\boldsymbol{A}\right)I]^{T}}{2Tr\left(\boldsymbol{A}\right)Det\left(\boldsymbol{A}\right)}\!.
\end{eqnarray}
Here, $Tr\left(\boldsymbol{A}\right)$ and $Det\left(\boldsymbol{A}\right)$
are the trace and the determinant of matrix $\boldsymbol{A}$, respectively.
We calculate the correlation matrices elements,
\begin{equation}
C_{11}=\frac{-2\kappa[\kappa^{\prime}+4\chi^{\prime\prime}n_{0}+4\zeta^{\prime}\Re(\alpha_{0})]^{\ast}(\chi^{\prime\prime}\alpha_{0}^{2}+\zeta^{\prime}\alpha_{0})(1+2n_{th})}{Tr\left(\boldsymbol{A}\right)Det\left(\boldsymbol{A}\right)},\label{eq:C11}
\end{equation}
 and
\begin{equation}
C_{12}=\frac{2\kappa n_{th}\vert\kappa^{\prime}+4\chi^{\prime\prime}n_{0}+4\zeta^{\prime}\Re\left(\alpha_{0}\right)\vert^{2}+4\kappa\vert\chi^{\prime\prime}\alpha_{0}^{2}+\zeta'\alpha_{0}\vert^{2}}{Tr\left(\boldsymbol{A}\right)Det\left(\boldsymbol{A}\right)}\label{eq:C12}
\end{equation}
 respectively. Then, we obtain the total photon number inside the
cavity including the quantum fluctuation effect as
\begin{eqnarray}
\bar{n} & = & n_{0}+C_{12}.\label{eq:total num}
\end{eqnarray}
 For the zero temperature case $n_{th}=0$, the above equation changes
into
\begin{align}
\bar{n} & =n_{0}+\frac{2\vert\chi_{kerr}\alpha_{0}^{2}+\zeta\alpha_{0}\vert^{2}}{Det\left(\boldsymbol{A}\right)}.\label{eq:red tot nm}
\end{align}
It follows from Eq. (\ref{eq:red tot nm}) that if $g=0$ the above
intercavity photon number is zero since all the parameters in our
system are proportional to the coupling coefficient $g$.

The second order correlation function can also be calculated easily
from the above correlation matrices elements as
\begin{eqnarray}
\!\!\!\!\! g^{\left(2\right)}\!\!\left(0\right) & \!\!\!\approx & \!\!\!1\!\!+\!2\!\frac{\langle\!\alpha_{1}^{\dagger}\alpha_{1}\!\rangle}{n_{0}}\!\!+\!2\!\Re\!(\!\frac{\langle\alpha_{1}^{2}\rangle}{\alpha_{0}^{2}}\!)\!\!=\!\!1\!\!+\!\frac{2C_{12}}{n_{0}}\!+\!2\Re\!(\!\frac{C_{11}}{\alpha_{0}^{2}}\!).\label{eq:SOCF}
\end{eqnarray}

\begin{figure}
\includegraphics[width=8cm]{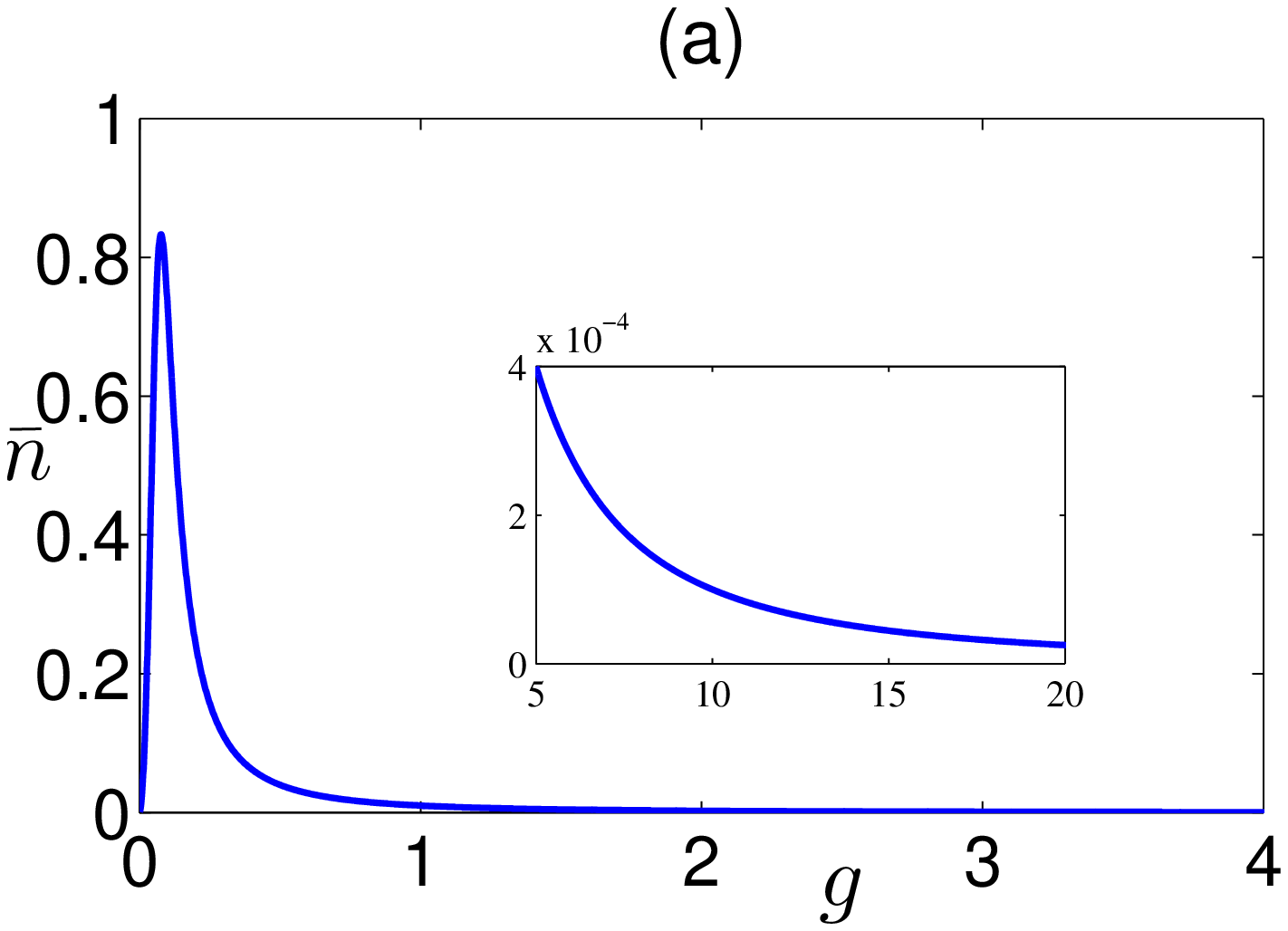}

\includegraphics[width=8cm]{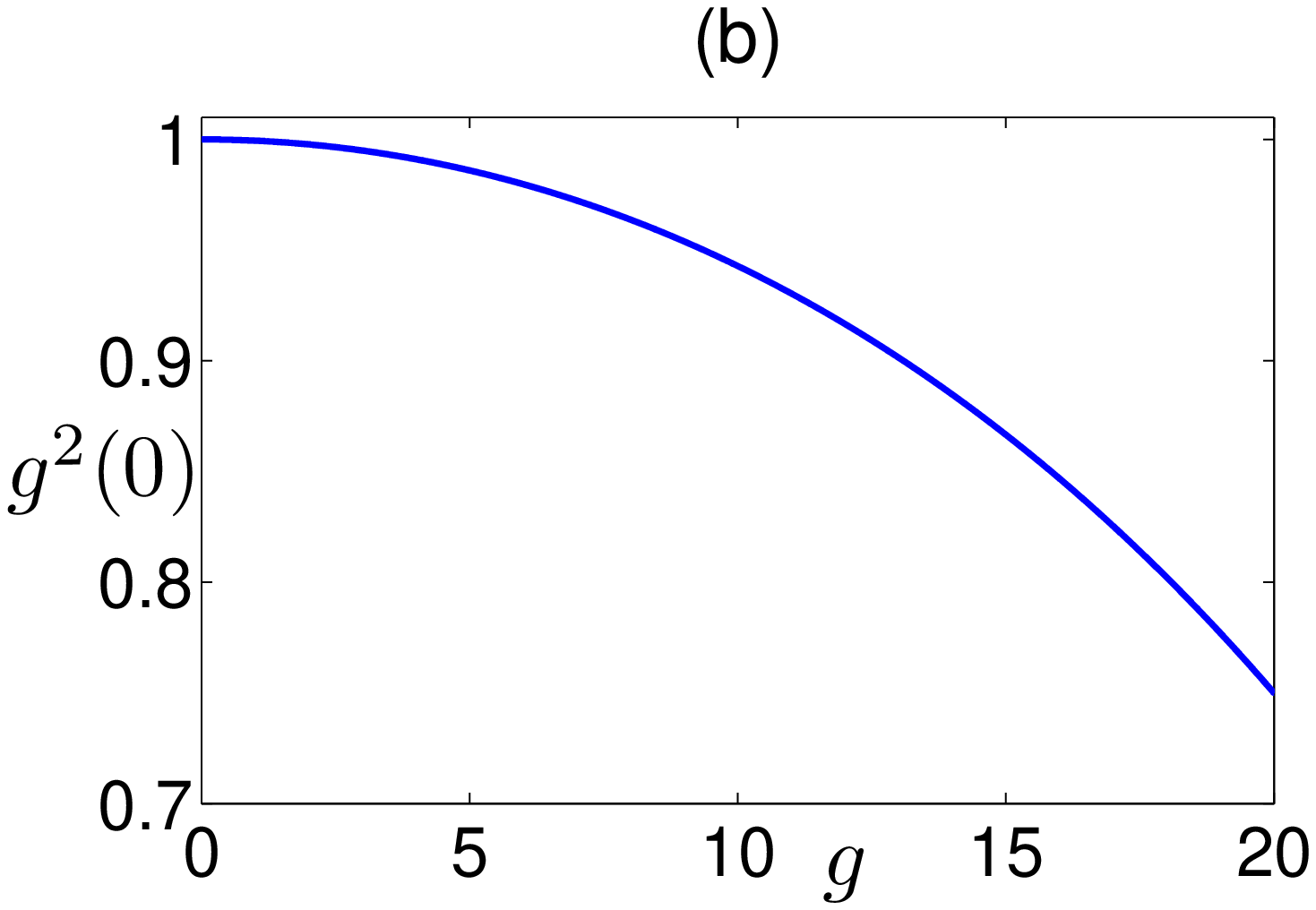}

\caption{\label{fig:SOCF}(Color online) The second order correlation function
$g^{\left(2\right)}\left(0\right)$, in Eq. $\left(\ref{eq:SOCF}\right)$
and total intracavity photon number in Eq. (\ref{eq:total num}) vs
coupling strength $g$ . Here take the parameters as : $\Delta_{c}=0$,
$\Delta_{b}=60$, $N=1\times10^{4}$, $\Omega=0.1$. All the parameters
are in the units of $\kappa$. }
\end{figure}

Generally, the thermal fluctuation would increase the second order
correlation function at zero time delay $g^{\left(2\right)}\left(0\right)$
to above unity. To optimize $g^{\left(2\right)}\left(0\right)$ to
investigate the photon antibunching effect of the system, we only
consider the zero temperature case, $n_{th}=0$. Thus the above second
order correlation Eq. (\ref{eq:SOCF}) is only related to the quantum
fluctuation effect. The intracavity photon number and second order
correlation function at zero time delay vs the coupling strength $g$
is shown in Fig. (\ref{fig:SOCF}). We just consider the case, where
the driving field is resonant with the cavity field $\Delta_{c}=0$
, but largely detunig from the atoms. As shown in Fig. \ref{fig:SOCF}(a),
the intracavity photon number will be much lower than one as increasing
the coupling strength $g$, and the corresponding $g^{\left(2\right)}\left(0\right)$
displays the typical antibunching behavior as shown in Fig. \ref{fig:SOCF}
(b). The strong nonlinear effects appears at very low photon number,
i.e, as low as $\bar{n}\simeq10^{-4}$.

\section{outpuit intensity and squeezing spectra }

In this section, we investigate the second extreme case $\Omega\gg g$,
where the squeezing effect of single-mode cavity field is dominant.
To calculate the output fluctuation spectrum, we write down the quantum
Langevin equation of cavity mode according to the Hamiltonian~(\ref{eq:Squeezing Hamil})
as

\begin{align}
\!\!\!\!\dot{c}\! & =\!-\!(\frac{\kappa}{2}\!+\! i\Delta_{eff,2})c\!-\!2i\mu c^{\dagger}\!-\!\! i\zeta\!(2c^{\dagger}c\!+\! c^{2})\!\!-\! iF^{\prime\prime}\!\!+\!\!\sqrt{\kappa}c_{in}\!.\label{eq:Sueezed lang}
\end{align}
Here, $c_{in}\left(t\right)$ is the noise operator andsatisfies the
fluctuation relations as listed in Eq. (\ref{eq: nfluc of c}). The
steady state value of $c$ is determined by
\begin{equation}
F^{\prime\prime}-i(\frac{\kappa}{2}+i\Delta_{eff,2})c_{s}+2\mu c_{s}^{\ast}+\zeta\left(2\vert c_{s}\vert^{2}+c_{s}^{2}\right)=0.\label{eq:ss of squ}
\end{equation}
To study the influence of the quantum fluctuation, we split the operator
$c$ into two parts $c=c_{s}+\delta c$. Here, $\delta c$ represents
the fluctuation operator, which has a vanishing mean value, i,e.,
$\langle\delta c\rangle=0$. Thus, after the linearization, the Langevin
equation (\ref{eq:Sueezed lang}) is rewritten as
\begin{align}
\!\!\!\!\delta\dot{c}\! & =\!-[\frac{\kappa}{2}\!+\! i\triangle_{eff,2}\!+\!4i\zeta\Re\!\left(\! c_{s}\!\right)]\delta c\!-\!2i\!\left(\!\mu\!+\!\zeta c_{s}\!\right)\!\delta c^{\dagger}\!+\!\sqrt{\kappa}c_{in}\!.\label{eq:fluc lang}
\end{align}
 By taking the Fourier transformation, we have
\begin{equation}
\delta c\left(\omega\right)=\frac{\sqrt{\kappa}}{D\left(\omega\right)}\left[-iBc_{in}^{\dagger}\left(\omega\right)+A^{*}\left(-\omega\right)c_{in}\left(\omega\right)\right].\label{eq:frou lan}
\end{equation}
 where\begin{subequations}
\begin{align}
A\left(\omega\right) & =-i\omega+[\frac{\kappa}{2}+i\Delta_{eff}^{\left(2\right)}+4i\zeta\Re(c_{s})],\label{eq: para A}\\
B & =2(\mu+\zeta c_{s}),\label{eq:para B}\\
D\left(\omega\right) & =A\left(\omega\right)A^{*}\left(-\omega\right)-\vert B\vert^{2}.\label{eq:para C}
\end{align}
\end{subequations}

\begin{figure}
\includegraphics[width=8cm]{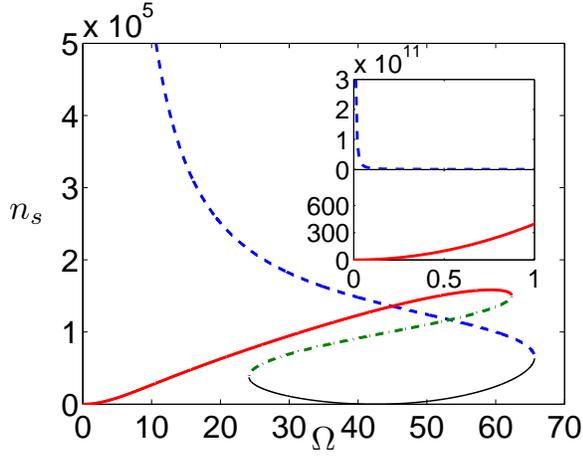}

\caption{(Color online) \label{fig:ph&int} Intracavity photon number vs the
driving strength. Here $N=1\times10^{4}$, $g=0.1$, $\Delta_{c}=1,$
$\Delta_{b}=60$. All parameters in the unit of the cavity decay rate,
$\kappa$. The different colors in this figure represents the different
steady-state values under same parameters, and the inset in this figure
depicts the solid red and dashed blue curves in the interval $\Omega\in(0,1]$. }
\end{figure}

In the frequency space, the noise operators satisfy the following
relations\begin{subequations}
\begin{eqnarray}
\langle c_{in}\left(\omega\right)c_{in}^{\dagger}\left(\omega'\right)\rangle & = & [n(\omega_{c})+1]\delta\left(\omega+\omega'\right),\label{eq:noise fluction rel1}\\
\langle c_{in}^{\dagger}\left(\omega\right)c_{in}\left(\omega'\right)\rangle & = & n\left(\omega_{c}\right)\delta\left(\omega+\omega'\right),\label{eq:noise fluctuation rel2}\\
\langle c_{in}\left(\omega\right)c_{in}\left(\omega'\right)\rangle & = & \langle c_{in}^{\dagger}\left(\omega\right)c_{in}^{\dagger}\left(\omega'\right)\rangle=0.\label{eq:noise fluctions rel3}
\end{eqnarray}
\end{subequations}The input-output relationship is given by $c_{out}=\sqrt{\kappa}c-c_{in}$.
After a linearization of the input-output fields around the steady
state value, the corresponding relationship between input and output
fluctuation operators in the frequency space reads as
\begin{equation}
\delta c_{out}\left(\omega\right)=\sqrt{\kappa}\delta c\left(\omega\right)-c_{in}\left(\omega\right).\label{eq:out put eq}
\end{equation}
The output intensity spectrum $S_{I}\left(\omega\right)$ \cite{Mancini}
is defined as
\begin{equation}
S_{I}\left(\omega\right)=\frac{1}{\vert c_{out}\vert^{2}}\int d\omega'\langle\delta I_{out}\left(\omega\right)\delta I_{out}\left(\omega'\right)\rangle,\label{eq:spectrum}
\end{equation}
where
\begin{equation}
\delta I_{out}\left(\omega\right)=c_{out}^{\ast}\delta c_{out}\left(\omega\right)+c_{out}\delta c_{out}^{\dagger}\left(\omega\right).\label{eq:out put op}
\end{equation}
By substituting Eq. (\ref{eq:frou lan}) and Eq. (\ref{eq:out put eq})
into Eq. (\ref{eq:spectrum}) and using the noise fluctuation relations
(\ref{eq:noise fluction rel1}-\ref{eq:noise fluctions rel3}), we
obtain the explicit expression of output intensity spectrum of single-mode
cavity field as
\begin{align}
S_{I}\left(\omega\right) & =\vert1-\frac{\kappa}{D\left(\omega\right)}\left(C\left(\omega\right)+ie^{2i\varphi}B^{\ast}\right)\vert^{2}.\label{eq:out spectrum}
\end{align}
Here, $\varphi$ is the phase of the output field, and its value is
determined by the input-output relationship. We note that inthe above
calculation the temperature $T$ of the cavity field is assumed to
be zero, i.e., $n\left(\omega_{c}\right)=0$.

\begin{figure}
\includegraphics[width=8cm]{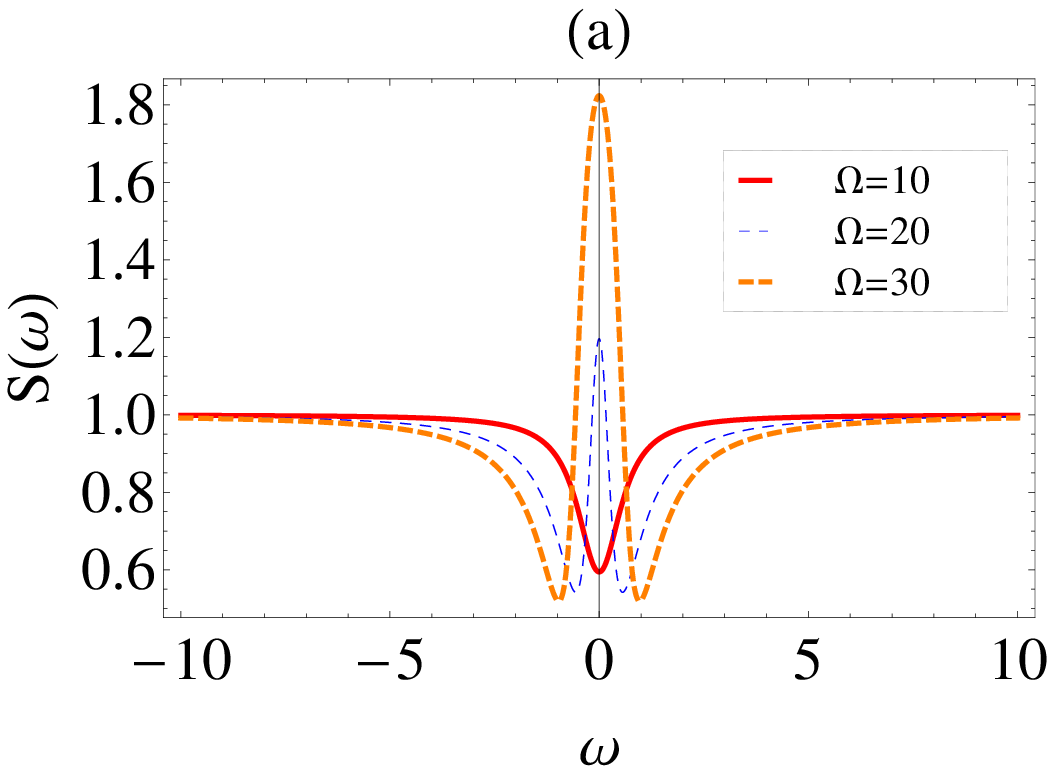}

\includegraphics[width=8cm]{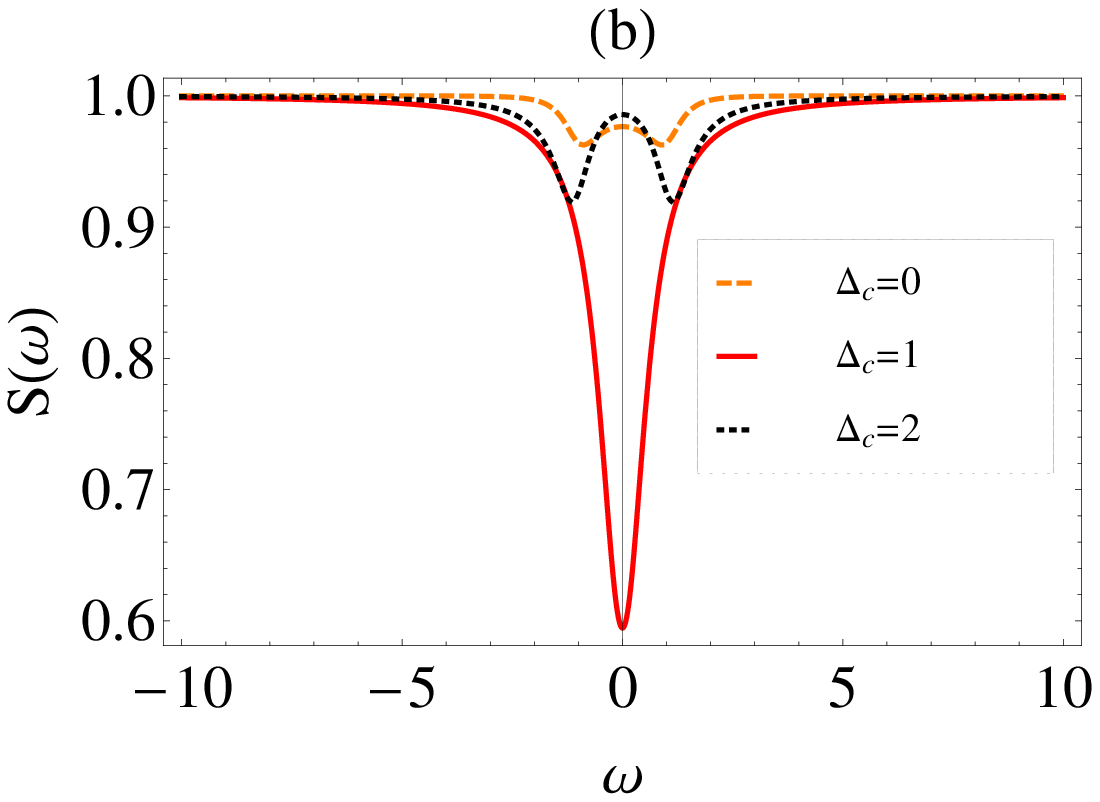}

\caption{(Color online) \label{fig:output spec}The output intensity spectrum
of single-mode cavity field for (a) different $\Omega$ with fixed
$\Delta_{c}=1$ and (b) different $\Delta_{c}$ with fixed $\Omega=10$.
Here, $N=1\times10^{4}$, $g=0.1$, and $\Delta_{b}=100$. All the
parameters are in the units of the cavity decay rate, $\kappa$. }
\end{figure}

The variation of the steady-state field intensity $n_{s}=\vert c_{s}\vert^{2}$,
which is determined by Eq. (\ref{eq:ss of squ}), as a function of
the driving field is given in Fig. \ref{fig:ph&int}. It is clear
that the bistability and even quad-stability would occur in our system
as increasing the driving field strength. The four lines with different
colors represent four different steady state solutions of Eq. (\ref{eq:ss of squ}).
As shown in the subplot of Fig. \ref{fig:ph&int}, the solid-red line
means $n_{s}$ starts from $0$ and increases with the driving strength
monotonically, and the dashed-blue line corresponds to that $n_{s}$
starts at infinite and decreases with the driving strength monotonically.
In the following, we choose the value of $n_{s}$ on the solid-red
line to do the forward calculations.

To investigate the squeezing effect of the system, we numerically
calculate the output intensity spectrum $S\left(\omega\right)$. As
shown in Fig. \ref{fig:output spec}(a), there would occur the squeezing
effect at detunig $\Delta_{c}=1$ and the single minimum peak is split
into two peaks when the the driving strength increases. From Fig.
\ref{fig:output spec}(b), we also find that the detuning between
the cavity mode and the driving frequency $\Delta_{c}$ also affects
the squeezing effect of the output intensity spectrum when the driving
strength is fixed.

\section{Conclusion and remarks}

In this paper we have studied the microscopic mechanism of the external
driving for a single-mode cavity field based on an indirect driving
model. In this simplified model a wall of the cavity is imagined as
an ensemble of local two-level systems. Through this modeling we investigated
the nonlinear effects of the single-mode cavity field, which is induced
by the re-combination of the higher-order excitations of atomic ensemble.
By adjusting some parameters there will occur the typical nonlinear
phenomena of the single-mode cavity field such as the photon blockade
and squeezing effects.

Our scheme in this paper is closely related to the microscopic description
of laser, and can be considered as a simplified non-linear quantum
optical model. Actually, generating the entangled photons is very
important in quantum information processing and quantum communication.
Our setup may provide a potential sources of entangled photons if
we consider the atomic ensemble with three-level configuration, which
can generate laser compared with the case of two-level configuration
\cite{Louseil} .
\begin{acknowledgments}
This work was supported by the National Natural Science Foundation
of China (Grant No.11121403, No.10935010 and No.11074261) and the
National 973 program (Grant No. 2012CB922104 and No. 2014CB921402).\end{acknowledgments}

\end{document}